%% file: ms.tex
\documentclass[conference]{IEEEtran}
\usepackage{algorithm,algorithmic}
\makeatletter
\newcommand\fs@betterruled{%
  \def\@fs@cfont{\bfseries}\let\@fs@capt\floatc@ruled
  \def\@fs@pre{\vspace*{5pt}\hrule height.8pt depth0pt \kern2pt}%
  \def\@fs@post{\kern2pt\hrule\relax}%
  \def\@fs@mid{\kern2pt\hrule\kern2pt}%
  \let\@fs@iftopcapt\iftrue}
\floatstyle{betterruled}
\restylefloat{algorithm}

\makeatother
\newcommand{\squeezeupan}{\vspace{-2mm}}

\usepackage{cite}
\usepackage{amsmath,amssymb,amsfonts}
\usepackage{algorithmic}
\usepackage{graphicx}
\usepackage{textcomp}
\usepackage{comment}
\usepackage{enumitem}
\usepackage{xcolor}
\def\BibTeX{{\rm B\kern-.05em{\sc i\kern-.025em b}\kern-.08em
    T\kern-.1667em\lower.7ex\hbox{E}\kern-.125emX}}
\input{./Definitions.tex}

\begin{document}
\title{Simultaneous Multi-User MIMO Communications and Multi-Target Tracking with Full Duplex Radios}

\author{\IEEEauthorblockN{Md Atiqul Islam\IEEEauthorrefmark{2}, George C. Alexandropoulos\IEEEauthorrefmark{4}, and Besma Smida\IEEEauthorrefmark{2}}
\IEEEauthorblockA{{\IEEEauthorrefmark{2}Department of Electrical and Computer Engineering, University of Illinois at Chicago, USA}\\
\IEEEauthorrefmark{4}Department of Informatics and Telecommunications, National and Kapodistrian University of Athens, Greece\\
emails: \{mislam23,smida\}@uic.edu, alexandg@di.uoa.gr
}}
\maketitle
\begin{abstract}
In this paper, we present an Integrated Sensing and Communications (ISAC) system enabled by in-band Full Duplex (FD) radios, where a massive Multiple-Input Multiple-Output (MIMO) base station equipped with hybrid Analog and Digital (A/D) beamformers is communicating with multiple DownLink (DL) users, and simultaneously estimates via the same signaling waveforms the Direction of Arrival (DoA) as well as the range of radar targets randomly distributed within its coverage area. Capitalizing on a recent reduced-complexity FD hybrid A/D beamforming architecture, we devise a joint radar target tracking and DL data transmission protocol. An optimization framework for the joint design of the massive A/D beamformers and the Self-Interference (SI) cancellation unit, with the dual objective of maximizing the radar tracking accuracy and DL communication performance, is presented. Our simulation results at millimeter wave frequencies using 5G NR wideband waveforms, showcase the accuracy of the radar target tracking performance of the proposed system, which simultaneously offers increased sum rate compared with benchmark schemes.
\end{abstract}


\section{Introduction}
Integrated Sensing and Communications (ISAC) is emerging as a key feature of the next-generation wireless networks, where sensing and communication signaling operations are unified in a single system to considerably improve spectral and energy efficiencies while reducing both hardware and signaling costs \cite{liu2022integrated,HRIS,liu2020joint,alexandropoulos2021hybrid_all,mishra2019toward}. In addition to its implementation in cellular networks, ISAC systems have recently been considered for a wide variety of applications, e.g., Wi-Fi networks \cite{liu2022integrated}, Unmanned Aerial Vehicle (UAV) networks \cite{wang2020constrained}, military communications \cite{liu2020joint}, and  localization for Vehicular networks (V2X)\cite{wymeersch20175g}. 
As a key enabler for ISAC applications, Full Duplex (FD) massive Multiple-Input Multiple-Output (MIMO) radios have the potential to be employed for the simultaneous DownLink (DL) transmission and UpLink (UL) reception capability within the entire frequency band \cite{Islam2019unified,alexandropoulos2020full,Islam_2020_Sim_Multi,sabharwal2014band}. FD multi-user massive MIMO systems in conjunction with fifth Generation (5G) millimeter Wave (mmWave) wideband waveforms can provide high-resolution radar target detection and tracking while ensuring high capacity communication links to DL users.

The principal bottleneck of the FD ISAC systems is the Self-Interference (SI) signal induced from the Transmitter (TX) to the Receiver (RX) at the massive MIMO FD Base Station (BS) node due to FD operation. 
Recently in \cite{sabharwal2014band,islam2021direction,Vishwanath_2020,alexandropoulos2020full}, a combination of propagation domain isolation, analog domain suppression, and digital SI cancellation techniques has been employed to achieve the required SI suppression for the mmWave FD massive MIMO transceivers. Hybrid Analog and Digital (A/D) BeamForming (HBF) is an attractive configuration for FD massive MIMO systems since it utilizes a small number of Radio Frequency (RF) chains connected to large-scale antenna arrays via phase shifters to reduce hardware cost. Appropriate A/D beamforming in the FD HBF system can reduce the impact of SI in the FD RX chains. Thus, a reduced complexity A/D SI cancellation solution can be formulated for FD massive MIMO systems with hybrid beamforming \cite{alexandropoulos2020full,Vishwanath_2020}.

Recently, single-antenna FD systems employing joint radar communication and sensing were introduced, where both communication and radar waveforms were studied for sensing performance \cite{barneto2019full,liyanaarachchi2021optimized}.
FD ISAC operations with mmWave massive MIMO systems were proposed in \cite{barneto2020beamforming,liyanaarachchi2021joint}. A multibeam approach with dedicated beams towards both a radar target and a DL user was considered in \cite{barneto2020beamforming}, whereas the authors in \cite{liyanaarachchi2021joint} provided an ISAC technique detecting the Direction of Arrival (DoA) of two radar targets, while only successfully estimating the range of one target. In \cite{Islam_2022_ISAC}, we proposed a reduced complexity single-user FD ISAC system with massive MIMO BS operating at mmWave frequencies capable of estimating both the DoA and range of multiple radar targets, while maximizing the DL rate. However, none of the previous works provide an FD ISAC massive MIMO system with radar target tracking protocols across multiple communication slots with simultaneous multi-user DL communication.

In this paper, we present a multi-user FD ISAC system including a protocol for multiple radar target DoA tracking and range estimation across several communication subframes. The considered ISAC system employs an FD massive MIMO BS node communicating with multiple DL users, and utilizes the reflected waveforms to detect and track the radar targets residing within the communication environment. We propose a joint design of the A/D beamformers and a reduced complexity SI cancellation for the FD ISAC system, which target at maximizing the multi-user DL communication rate and the precision of the radar target tracking. We perform an extensive waveform simulation with 5G wideband Orthogonal Frequency Division Multiplexing (OFDM) waveforms at mmWave frequencies, verifying the performance of the proposed multi-user FD ISAC system. 

\section{System and Signal Models}
\begin{figure}[!tpb]
	\begin{center}
	\includegraphics[width=\linewidth]{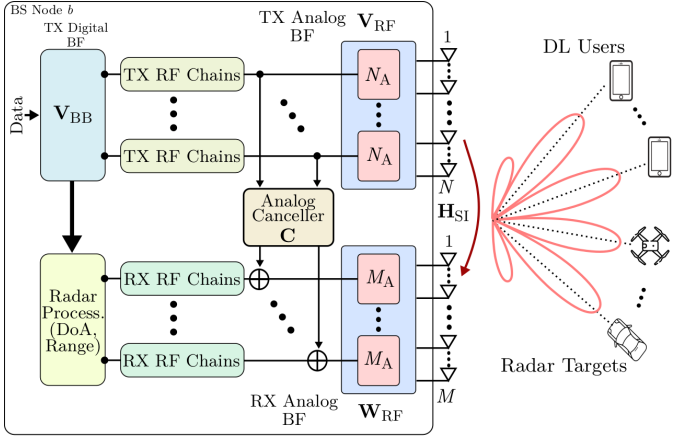}
	\caption{The considered FD-enabled massive MIMO ISAC system, where the FD hybrid A/D beamforming BS communicates with DL users and simultaneously detects corresponding DoA and range of radar targets.}
	\label{fig: FD_ISAC}
	\end{center}
	\squeezeupan
\end{figure}

We consider a multi-user FD massive MIMO ISAC system operating at mmWave frequencies, where an FD massive MIMO BS node is communicating with $U$ RX user nodes in the DL direction, as depicted in Fig. \ref{fig: FD_ISAC}. The DL signals are reflected by the multiple radar targets distributed within the communication environment, which are received and processed at the RX of BS node for radar targets' parameter estimation enabling integrated sensing and communication. 

The FD massive MIMO BS node $b$ is comprised of $N$ TX and $M$ RX antennas, whereas each of the $U$ users has $L$ RX antennas. To reduce the hardware complexity in massive MIMO BS node, we consider a small number of TX/RX RF chains partially-connected to Uniform Linear Arrays (ULAs) of large number of antenna elements via analog phase shifters following a Hybrid BeamForming (HBF) structure. Therefore, in the BS node, each of the $N_{\rm RF}$ and $M_{\rm RF}$ TX/RX RF chains are connected to ULAs of $N_{\rm A}$ and $M_{\rm A}$ antenna elements, respectively. The configurations of the phase shifters are contained in analog beamformers $\V_{\mathrm{RF}}={\rm diag}(\v_1,\dots,\v_{N_{\rm RF}})\in\Compl^{N\times N_{\rm RF}}$ and $\W_{\mathrm{RF}}={\rm diag}(\w_1,\dots,\w_{N_{\rm RF}})\in\Compl^{M \times M_{\rm RF}}$, respectively. The elements of the TX/RX analog BFs are assumed to have constant magnitude and chosen from predefined beam codebooks, i.e., $\v_n \in \mathbb{F}_{\rm TX}\, \forall n= 1,\dots,N_{\rm RF}$ and $\w_m \in \mathbb{F}_{\rm RX}\, \forall m= 1,\dots,M_{\rm RF}$. The TX/RX beam codebooks consists of ${\rm card}(\mathbb{F}_{\rm TX})$ and ${\rm card}(\mathbb{F}_{\rm RX})$ distinct analog beams, respectively. The RX user nodes are considered to employ fully digital beamforming, since the number of the user antennas is typically much smaller than at the FD massive MIMO BS.

\subsection{DL and Radar Reflected Signals}
We assume a 5G NR subframe-based DL signaling operation for the considered multi-user FD massive MIMO ISAC system. In each subframe, the BS transmits mmWave 5G NR OFDM waveforms to the DL users comprising $Q$ OFDM symbols with $P$ active subcarriers and $\Delta f$ subcarrier spacing. In addition to the DL communication, these OFDM symbols are reflected by multiple radar targets and received at the BS RX, which is utilized for tracking targets across subframes.

To enable multi-user MIMO communication, each subcarrier of the DL waveform contains $L$ parallel data streams for each of the $U$ DL users such that $UL\leq N_{\rm RF}$. In the BaseBand (BB), the $u$th user's unit frequency-domain symbol vector $\s_{p,q,u}\in\Compl^{L}$ at the $p$th subcarrier of $q$th OFDM symbol is precoded using digital beamforming matrix $\V_{{\rm BB},u}\in\Compl^{N_{\rm RF}\times L}\, \forall u = 1,\dots,U$. Furthermore, the precoded signals are processed by the analog BF and the transmitted frequency-domain symbol vector $\x_{p,q}\in\Compl^{N}$ at the antenna elements can be written as
\begin{equation}\label{eq:TX_Signal}
\begin{split}
   \x_{p,q} &\triangleq \V_{\mathrm{RF}}[\V_{\mathrm{BB},1},\dots\V_{\mathrm{BB},U}][\s_{p,q,1}^{\rm T},\dots,\s_{p,q,U}^{\rm T}]^{\rm T}\\
& = \V_{\mathrm{RF}}\V_{\mathrm{BB}}\s_{p,q},
\end{split}
\end{equation}
where $\V_{\mathrm{BB}} \triangleq [\V_{\mathrm{BB},1},\dots\V_{\mathrm{BB},U}]\in\Compl^{N_{\mathrm{RF}}\times UL}$ and $\s_{p,q} \triangleq [\s_{p,q,1}^{\rm T},\dots,\s_{p,q,U}^{\rm T}]^{\rm T}\in\Compl^{UL}$. The DL transmission is power limited such that $\mathbb{E}\{\|\V_{\mathrm{RF}}\V_{\mathrm{BB}}\s_{p,q}\|^2\}\leq {\rm P}_b$, where ${\rm P}_b$ represents the maximum transmission power at node $b$.

For the integrated sensing operation, we assume $K$ radar targets/scatters randomly distributed within the communication/sensing environment. Each of the $K$ targets is associated with a DoD/DoA\footnote{It is to be noted that direction of departure and arrival of radar targets are identical since we consider a monostatic radar setup assuming relatively far away targets and small TX-RX array separation.} $\theta_k \in [-\pi\,\,\pi]$ and a range $\delta_k$ from the BS node corresponding to a respective delay $\tau_k \triangleq 2\delta_k/c$, where $c$ represents the speed of light. These radar targets reflect the DL transmitted signal $\x_{p,q}$, which is received at the RX of the FD BS node $b$. The radar RX signal $\y_{p,q}\in\mathbb{C}^{M}$ comprising reflected and SI signals is expressed as
\begin{equation}\label{eq:Radar_Signal}
    \begin{split}
        \y_{p,q} \triangleq & \sum\limits_{k=1}^{K} \alpha_k e^{-j2\pi \tau_k p\Delta f} \a_{M}(\theta_k)\a_{N}^{\rm H}(\theta_k) \x_{p,q} 
        \\&+ \H_{\rm SI} \x_{p,q} + \n_{p,q},
    \end{split}
\end{equation}
where $\n_{p,q}\sim \mathcal{CN}(0,\sigma_b^2\I_{M})$ and $\alpha_k\in\Compl$ denote the receiver noise vector with covariance $\sigma_b^2$ and the reflection coefficient of the $k$th radar target, respectively. The propagation delay $\tau_k$ induces the phase shift $e^{-j2\pi\tau_k p \Delta f}$ across subcarriers \cite{braun2010maximum}. Here, the $i$th element of the ULA response vector $\a_{N}(\theta)\in\mathbb{C}^N $ with $N$ antenna elements and any DoA $\theta$ is expressed as
\begin{equation}
    \begin{split}
        [\a_{N}(\theta)]_{i} \triangleq  \frac{1}{\sqrt{N}}e^{-j\frac{2\pi}{\lambda}(i-1)d\sin(\theta)},
    \end{split}
\end{equation}
where $\lambda$ and $d$ are the signal wavelength and the inter-antenna element distance, respectively.
Here, $\H_{\rm SI}\in\Compl^{M\times N}$ is the Line-of-Sight (LoS) SI channel path between the TX and RX antenna arryas of the BS node $b$, which can be modeled as
\begin{equation}\label{eq: SI_channel}
    \begin{split}
        [\H_{\rm SI}]_{(m,n)} \triangleq \frac{\rho}{r_{m,n}}e^{-j\frac{2\pi}{\lambda}r_{m,n}},
    \end{split}
\end{equation}
where, $\rho$ represents the power normalization constant such that $\mathbb{E}\{\|[\H_{\rm SI}]_{(m,n)}\|_{F}^2\}= MN$. Here, $r_{m,n}$ denotes the distance between $m$th RX and $n$th TX antenna elements at the BS node, which depends on the transmit and receive array geometry \cite[eq. (9)]{satyanarayana2018hybrid}.

The received signal at the node $b$ RX is processed by the RX analog combiner $\W_{\rm RF}$, which is followed by analog and digital cancellation to suppress the LoS SI signal below the noise floor. In the RX BB of the BS, the frequency-domain radar reflected symbol vector $\widetilde{\y}_{p,q} \in \Compl^{M_{\rm RF}}$ can be expressed as
\begin{align}
        \nonumber\widetilde{\y}_{p,q} \triangleq &  \W_{\rm RF}^{\rm H}\!\sum\limits_{k=1}^{K} \!\!\alpha_k e^{-j2\pi \tau_k p\Delta f} \a_{M}\!(\theta_k)\a_{N}^{\rm H}\!(\theta_k) \x_{p,q}\\
        &+  (\widetilde{\H}_{\rm SI} + \C + \D)\V_{\rm BB}\s_{p,q} + \W_{\rm RF}^{\rm H}\n_{p,q},
\end{align}
where $\C\in \Compl^{M_{\rm RF}\times N_{\rm RF}}$ and $\D \in \Compl^{M_{\rm RF}\times N_{\rm RF}}$ represent the low complexity analog and digital SI cancellers, respectively. After the analog cancellation, the residual SI signal satisfy the RX RF saturation constraint, i.e., $\|[(\widetilde{\H}_{\rm SI} \!+\! \C)\V_{\rm BB}]_{(m,:)}\|^2\leq \rho_b,\forall m=1,\dots,M_{\rm RF}$, where $\rho_b$ represents the saturation level of RX RF chains at the BS node $b$. It is to be noted that the low complexity analog canceller $\C$ suppressing the LoS SI components is designed following the similar structure in \cite{alexandropoulos2020full}. Here, $\widetilde{\H}_{\rm SI}\triangleq\W_{\rm RF}^{\rm H}\H_{\rm SI}\V_{\rm RF}$ represents the effective LoS SI channel after analog TX/RX beamforming.

For the multi-user DL communication, we assume $U$ out of $K$ scatterers contribute to the DL channels from the BS node $b$ to the $U$ users. For each scatter, $\theta_u, \forall u = 1,\dots,U$ represents the DoD, while $\phi_u$ denotes the DoA at the user node. The received DL signal vector at the $u$th user $\r_{p,q,u}\in\Compl^{L}$ is expressed as
\begin{align}
        \nonumber\r_{p,q,u} &\triangleq \beta_{u}  \a_{L}(\phi_{u})\a_{N}^{\rm H}(\theta_{u})\x_{p,q} + \z_{p,q,u}\\ &=\mathbf{H}_{\mathrm{DL},u}\x_{p,q}+ \z_{p,q,u},
\end{align}
where $\beta_{u}\in\Compl$ and $\z_{p,q,u}\sim \mathcal{CN}(0,\sigma_u^2\I_{L})$ represent reflection coefficient of $u$th scatter and the noise floor at RX node $u$ with covariance $\sigma_u^2$, respectively. Here, $\mathbf{H}_{\mathrm{DL},u}\triangleq \beta_{u}  \a_{L}(\phi_{u})\a_{N}^{\rm H}(\theta_{u})$ represents the DL channel from BS node to $u$th user.

\section{FD-Enabled Multi-Target Tracking}
In this section, we present the proposed FD ISAC DL data transmission and multiple radar target estimation and tracking operation. We utilize the received reflected signals at the BS node RX for estimation and tracking.

\subsection{Radar Targets/Scatterers Evolution}
We assume a 5G NR subframe-based DL communication system, where each radio subframe of $T_s$ duration contains $Q$ OFDM symbols with $P$ subcarriers. The radar targets' and communication scatters' parameters are considered to remain constant for one subframe, while the parameters of the successive subframes are temporarily correlated. For any consecutive $(i\!-1\!)$ and $i$th subframes, the evolution of radar DoA components is expressed similar to \cite{va2016beam} as
\begin{equation}
    \begin{split}
        \theta_{k}[i] \triangleq \theta_{k}[i-1] + \Delta\theta_{k},\quad \forall k =1,\dots,K,
    \end{split}
\end{equation}
where $\Delta\theta_{k}$ depends on the velocity of the $k$th radar target and the subframe duration $T_s$. For simplicity, we assume that all the radar targets are moving with a constant velocity $v$ in a circular direction from the BS, hence, $\Delta\theta_{k} \triangleq \text{arctan}\left(\frac{vT_s}{\delta_{k}}\right)$. For brevity, the radar targets' complex-valued reflection coefficients $\alpha_k$ and DL channels complex path gains $\beta_u,\forall u$ are assumed to change randomly between consecutive time slots. In this paper, we propose to track only the radar targets' DoA components for each time subframe. The estimation of all complex path gains and reflection coefficients is left for future investigation.

\begin{figure}
    \centering
    \includegraphics[width=\linewidth]{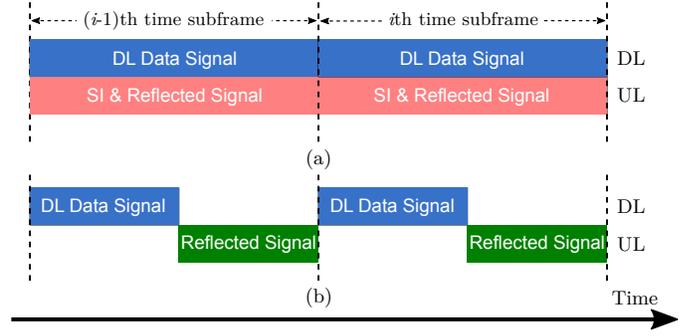}
    \caption{(a) The Proposed FD-enabled ISAC scheme and (b) the conventional HD-based ISAC signaling in each time subframe.}
    \label{fig: trans_protocol}
\end{figure}

\subsection{ISAC Multi-Target Tracking Operation Protocol}
The proposed FD ISAC protocol for DL data transmission and multiple radar target estimation and tracking is illustrated in Fig.~\ref{fig: trans_protocol}(a), where the DL channel is dedicated for data transmission to the users while the UL is accessed at the BS node to receive the reflected signals from radar targets simultaneously. The procedures of FD ISAC transmission are described as follows:
\begin{enumerate}
    \item At any $(i\!-\!1)$th time subframe, the DL signal reflected by the radar targets is received by the FD BS node $b$ enabled by reduced complexity massive MIMO FD analog and digital canceller.
    \item Utilizing the received signal after SI suppression, the DoAs of all $K$ targets during $(i\!-\!1)$th time subframe $\widehat{\theta}_{k}[i\!-\!1],\forall k$ are estimated at the BS node $b$ along with their range $\widehat{\delta}_k,\forall k$.
    \item During the DL data transmission to $U$ users at $i$th time subframe, the estimated DoAs $\widehat{\theta}_{k}[i\!-\!1]$ are used to choose appropriate digital beamformer $\V_{\mathrm{BB}}[i]$ and TX/RX phase shifter configurations $\V_{\mathrm{RF}}[i]$ and $\W_{\mathrm{RF}}[i]$ to assure maximized DL rate and satisfy RF saturation level constraint at the BS node $b$.
\end{enumerate}
In Fig.~\ref{fig: trans_protocol}(b), a conventional HD ISAC system is illustrated, where, contrary to the FD case, a portion of the subframe is dedicated for DL transmission while the rest is utilized for radar target detection.

\subsection{Multiple Radar Targets' DoA Estimation}
At the $(i\!-\!1)$th time subframe, we utilize the received reflected signal $\widetilde{\y}_{p,q}[i\!-\!1]$ to estimate the DoAs of $K$ radar targets employing MUltiple SIgnal Classification (MUSIC) algorithm. First, we calculate the sample covariance matrix across all subcarriers and OFDM symbols of $(i\!-\!1)$th subframe as
\begin{equation}\label{eq: sampl_cov}
    \begin{split}
         \widehat{\R}_b[i\!-\!1]\triangleq\frac{1}{PQ}\sum\limits_{q=0}^{Q-1}\sum\limits_{p=0}^{P-1}\widetilde{\y}_{p,q}[i\!-\!1]\widetilde{\y}_{p,q}^{\rm H}[i\!-\!1].
    \end{split}
\end{equation}
Now, we perform the eigenvalue decomposition of the estimated sample covariance matrix $\widehat{\R}_b[i\!-\!1]$ as
\begin{equation}\label{eq: cov_eig}
    \begin{split}
        \widehat{\R}_b[i\!-\!1] \triangleq \U {\rm diag}\{\eta_1,\eta_2,\ldots,\eta_{M_{\rm RF}}\}\U^{\rm H},
    \end{split}
\end{equation}
where $\eta_1\geq\eta_2\geq\ldots\geq\eta_{M_{\rm RF}}$ are the eigenvalues in the descending order and $\U\in\mathbb{C}^{M_{\rm RF}\times M_{\rm RF}}$ represents the eigenvector matrix. The matrix $\U$ is partitioned into signal and noise subspace as $\U=[\U_s|\U_n]$, where $\U_{n}\in\mathbb{C}^{M_{\rm RF}\times M_{\rm RF}-K}$ and $\U_s\in\mathbb{C}^{M_{\rm RF}\times K}$ contains the noise and signal subspace eigenvectors. Now, we estimate the DoAs of $K$ radar targets finding $K$ peaks of the MUSIC spectrum formulated as
\begin{equation}\label{eq: spectral_peak}
    \begin{split}
        P_{\rm MUSIC}(\theta) \triangleq \frac{1}{\a_{M}^{\rm H}(\theta)\W_{\rm RF}[i\!-\!1]\U_{n}\U_{n}^{\rm H}\W_{\rm RF}^{\rm H}[i\!-\!1]\a_{M}(\theta)}.
    \end{split}
\end{equation}
The $K$ peaks of $P_{\rm MUSIC}(\theta)$ correspond to the $K$ estimated DoAs $\widehat{\theta}_k[i\!-\!1],\forall k$.

\subsection{Multiple Radar Targets' Range Estimation}
Now, we estimate the range of $K$ radar targets at the $(i\!-\!1)$th subframe. As mentioned before, the range of radar targets $\delta_k,\forall k$ corresponds to propagation delay $\tau_k,\forall k$. Here, we obtain the estimated delay at $(i\!-\!1)$th subframe $\tau_k[i\!-\!1],\forall k$ from which respective range can be achieved. Utilizing the estimated DoAs $\widehat{\theta}_k[i\!-\!1]$ and the BB transmit signal vector $\x_{p,q}[i\!-\!1]$, we formulate a reference signal in the radar target direction as $\g_{p,q} \triangleq \mathbf{a}_{M}(\widehat{\theta}_k[i\!-\!1])\mathbf{a}_{N}^{\rm H}(\widehat{\theta}_k[i\!-\!1])\x_{p,q}[i\!-\!1]$. Now, we formulate a quotient across all RX antennas which includes the propagation delay impact in the direction of $\widehat{\theta}_k[i\!-\!1]$ as
\begin{equation}
    \begin{split}
        z_{p,q} \triangleq \frac{1}{M}\sum\limits_{m=1}^{M}\frac{[\W_{\rm RF}[i\!-\!1]\widetilde{\mathbf{y}}_{p,q}[i\!-\!1]]_m}{[\g_{p,q}[i\!-\!1]]_m},\,\,\forall p,q,
    \end{split}
\end{equation}
The quotient $z_{p,q}$ is utilized to formulate a  likelihood function as $A(n) \triangleq \sum\limits_{q=0}^{Q-1}\sum\limits_{p=0}^{P-1} z_{p,q} e^{j2\pi\frac{pn}{P}}$, where $n = 0,\dots,P-1$ is the quantized delay parameters. Now, we find the best quantized delay that maximizes the likelihood function norm as follows
\begin{equation}
    \begin{split}
        {n}^*\triangleq \underset{n}{\text{arg max}}\quad |A(n)|^2.
    \end{split}
\end{equation}
Therefore, estimated delay of the $k$th target at $(i\!-\!1)$th subframe is expressed as $\widehat{\tau}_k[i\!-\!1] \triangleq \frac{{n}^*}{P\Delta f}$ and the range is formulated as $\widehat{\delta}_k [i\!-\!1] \triangleq \frac{\widehat{\tau}_k[i\!-\!1] c}{2}$.
The DoA and delay estimation technique is utilized for any subframe to track the radar targets distributed within the communication environment.
\begin{algorithm}[!t]
    \caption{Multi-user FD-Based ISAC Optimization}
    \label{alg:the_opt}
    \begin{algorithmic}[1]
        \renewcommand{\algorithmicrequire}{\textbf{Input:}}
        \renewcommand{\algorithmicensure}{\textbf{Output:}}
        \REQUIRE $\widehat{\H}_{\rm SI}[i\!-\!1]$, $\widehat{\H}_{\mathrm{DL},u}[i\!-\!1]$, $N_{C}$, ${\rm P}_b$ and $\widehat{\theta}_k[i\!-\!1]\,\,\forall k$. 
        \ENSURE $\V_{\rm RF}[i],\V_{\rm BB}[i], \W_{\rm RF}[i],\C[i]$, and $\D[i]$.
        \STATE Set $\widehat{\H}_{\rm R}[i]\! =\! \sum\limits_{k=1}^{K}\! \mathbf{a}_{M}(\widehat{\theta}_k[i\!-\!1])\mathbf{a}_{N}^{\rm H}(\widehat{\theta}_k[i\!-\!1])$.
        \STATE Set $\V_{\rm RF}[i] \triangleq \underset{\v_n\in\mathbb{F}_{\rm TX},\forall n}{\text{arg max}}\, \|\widehat{\H}_{\rm R}[i] \V\|^2$.
        \STATE Set $\W_{\rm RF}[i] \triangleq \underset{\w_m\in\mathbb{F}_{\rm RX},\forall m}{\text{arg max}}\, \frac{\|\W^{\rm H}\widehat{\H}_{\rm R}[i] \V_{\rm RF}[i]\|^2}{\|\W^{\rm H}\widehat{\H}_{\rm SI}[i\!-\!1] \V_{\rm RF}[i]\|^2}$.
        \STATE Set $\widehat{\widetilde{\H}}_{\rm SI}[i] = \W_{\rm RF}^{\rm H}[i]\widehat{\H}_{\rm SI}[i\!-\!1]\V_{\rm RF}[i]$, $\widehat{\widetilde{\H}}_{\mathrm{DL},u}[i] = \widehat{\H}_{\mathrm{DL},u}[i\!-\!1]\V_{\rm RF}[i],\forall u$.
        \STATE Set $\C[i]$, $\D[i]$ as in \cite[Alg. 2]{Islam_2022_ISAC}, and $\B$ as the $N_{\rm RF}$ right-singular vectors of $(\widehat{\widetilde{\H}}_{\rm SI}[i]\!+\!\C[i])$.
        \FOR{$\alpha={N}_{\rm RF},N_{\rm RF}-1,\ldots,2$}
    		\STATE Set $\F\!=\![\B]_{(:,N_{\rm RF}-\alpha+1:N_{\rm RF})}$, $\H_{{\rm eff},u}\!=\!\widehat{\widetilde{\H}}_{\mathrm{DL},u}[i]\F,\forall u$.
    		\STATE Set $\bar{\E}_u\in \Compl^{N_{\rm RF}\times L}$ as the null space of effective DL channel with $u$th user removed $\bar{\H}_{{\rm eff},u}\triangleq[\H_{{\rm eff},1}^{\rm T},\ldots,\H_{{\rm eff},u-1}^{\rm T},\H_{{\rm eff},u+1}^{\rm T},\ldots,\H_{{\rm eff},U}^{\rm T}]$.
    		\STATE Set ${\E}_u$ as the right singular vectors of $\H_{{\rm eff},u}\bar{\E}_u$.
    		\STATE Set $\G_u = \sqrt{{{\rm P}_b/U}}\bar{\E}_u{\E}_u$ as the optimum block-diagonalized precoder for $u$th user given ${\rm P}_b$.
    		\STATE Obtain digital precoder $\G=[\G_1,\dots,\G_U]$ repeating steps $8$-$10$ for all $U$ users.
    		\IF{$\big\|[(\widehat{\widetilde{\H}}_{\rm SI}[i]\! +\! \C[i])\F\G]_{(m,:)}\big\|^2 \!\!\!\!\leq \! \rho_b , \!\forall m$,}
    			 \STATE Output $\V_{\mathrm{BB}}[i]=\F\G$ and stop the algorithm.
    	    \ELSE
            			 \STATE Output $\C[i]$ does not meet 
            			 the residual SI constraint.
    		\ENDIF
    	\ENDFOR
    \end{algorithmic}
\end{algorithm}
\section{Proposed Optimization Framework}
In this section, we focus on the joint design of the A/D beamformers $\mathbf{V}_{\rm BB}[i], \mathbf{V}_{\rm RF}[i], \mathbf{W}_{\rm RF}[i]$ and SI cancellation matrices $\C[i],\D[i]$ at the $i$th subframe to optimize multi-user MIMO communication and multiple radar target tracking performance.

Our proposed FD ISAC optimization framework maximizes the Signal-to-Noise-Ratio (SNR) to the DL users as well as all the radar targets to optimize both tracking and DL communication. Utilizing the estimated DoAs of $(i\!-\!1)$th subframe $\widehat{\theta}_k[i\!-\!1],\,\forall k$, the SNR in all the radar targets' direction at the $i$th subframe can be expressed as,
\begin{align}
        \nonumber\widehat{{\Gamma}}_{\rm R}[i] \triangleq \Big\|\W_{\rm RF}^{\rm H}[i]\sum\limits_{k=1}^{K}&\a_{M}(\widehat{\theta}_k[i\!-\!1])\a_{N}^{\rm H}(\widehat{\theta}_k[i\!-\!1])\V_{\rm RF}[i]\\
        &\times\V_{\rm BB}[i]\Big\|^2 \widehat{\Sigma}_b^{-1}[i],
\end{align}
where $\widehat{\Sigma}_b[i] \!\!=\!\! \|(\widehat{\widetilde{\H}}_{\rm SI}[i] \!+\! \C[i]\!+\!\D[i])\V_{\rm BB}[i]\|^2 \!+\! \|\W_{\rm RF}[i]\|^2\sigma_b^2$ represents the residual SI plus noise covariance. With the objective to maximize DL rate, we devise the sum of $U$ DL users' SNR as
\begin{equation}
    \begin{split}
        \widehat{{\Gamma}}_{\rm DL}[i]\!\triangleq\! \sum\limits_{u=1}^{U}\! \Big\|\widehat{\H}_{\mathrm{DL},u}[i\!\!-\!\!1]\V_{\rm RF}[i]\V_{\rm BB}[i]\Big\|^2\!\! \sigma_u^{-2}.
    \end{split}
\end{equation}
We formulate the optimization problem maximizing both SNR in the Radar target direction and DL users as
\begin{align}\label{eq: optimization_eq}
        \mathcal{OP}&:\nonumber\underset{\substack{\V_{\rm RF}[i],\V_{\rm BB}[i],\W_{\rm RF}[i]\\ \C[i],\D[i]}}{\max} \quad \widehat{ {\Gamma}}_{\rm R}[i]+ \widehat{ {\Gamma}}_{\rm DL}[i]\\
        &\text{\text{s}.\text{t}.}\,
        \big\|[(\widehat{\widetilde{\H}}_{\rm SI}[i]\! +\! \C[i])\V_{\rm BB}[i]]_{(m,:)}\big\|^2 \!\!\!\!\leq \! \rho_b , \!\forall m\! = 1, \ldots, M_{\rm RF},\nonumber\\
        &\,\quad\mathbb{E}\{\|\V_{\rm RF}[i]\V_{\rm BB}[i]\|^2\}\leq {\rm P}_b,\\
        &\,\quad\mathbf{w}_m[i]\in\mathbb{F}_{\rm RX},\forall m\,\,{\rm and}\,\, \mathbf{v}_n[i]\in\mathbb{F}_{\rm TX},\forall n=1,\ldots,N_{{\rm RF}}\nonumber
\end{align}
We employ an alternating optimization approach to solve this non-convex problem suboptimally. First, using the estimated DoAs $\widehat{\theta}_k[i\!-\!1]\,\,\forall k$ at any $(i\!-\!1)$th subframe, we find for the analog TX/RX BFs that maximizes the signal power toward all radar direction while minimizing the SI channel impact at the RX of BS node $b$. Utilizing the analog BFs, we follow a similar procedure as in \cite{Islam_2022_ISAC} to find the $N_{C}\leq {N}_{\rm RF}{M}_{\rm RF}$ taps analog canceller.
Finally, we design the multi-user digital precoder $\V_{\rm BB}$ using block diagonalization such that it maximizes the SNR of the DL users while minimizing the inter-user interference and suppressing the residual SI at node $b$. The proposed solution for \eqref{eq: optimization_eq} is summarized in Algorithm \ref{alg:the_opt}.

\begin{table}[tbp]
    \caption{Waveform and System Level Parameters}
    \label{tab:1}
    \centering
    \footnotesize
    \begin{tabular}{|c||c|}
         \hline
         \!\!\!\textbf{Parameter}\!\!\! & \!\!\!\textbf{Value}\!\!\! \\
         \hline
        \!\!\! Carrier Frequency\!\!\! &\!\!\! $28$ GHz\!\!\!\\
         \!\!\!Bandwidth\!\!\! & \!\!\!$100$\!\!\! MHz\!\!\! \\
         \!\!\!Subframe Duration, $T_s$\!\!\! & \!\!\!$1$ms\!\!\! \\
         \!\!\!OFDM Symbols, $Q$\!\!\! &\!\!\! $14$\!\!\!\\
         \!\!\!Data Subcarriers, $P$\!\!\! & \!\!\! $792$\!\!\!\\
         \!\!\!Subcarrier Spacing, $\Delta f$\!\!\! & \!\!\!$120$ KHz\!\!\!\\
         \hline
    \end{tabular}
    \begin{tabular}{|c ||c|}
         \hline
         \textbf{Parameter} & \textbf{Value} \\
         \hline
         Transmit Power & $10$:$30$ dBm\\
         RX Noise Floor & $-90$ dBm\\
         ADC Bits & $14$bits\\
         PAPR & $10$ dB\\
         Dynamic Range & $60$ dB\\
         RF Saturation & $-30$ dBm\\
         \hline
    \end{tabular}
\end{table}

\section{Numerical Results}
In this section, we present the numerical evaluation of the proposed FD massive MIMO multi-user communication and multi-target tracking system through extensive waveform simulation.

\subsection{Simulation Parameters}
Following the FD massive MIMO architecture in Fig.~\ref{fig: FD_ISAC}, we consider a $128\times 128$ FD massive MIMO BS node $b$ with $N_{\rm RF} = M_{\rm RF} = 8$ TX/RX RF chains. Since the BS node employs a partially-connected beamforming structure, each TX/RX RF chain is connected to a ULA of $N_{\rm A} = M_{\rm A} = 16$ antenna elements via phase shifters. Both TX/RX phase shifter configurations (analog beams) are chosen from $5$-bit Discrete Fourier Transform (DFT) codebooks. The multi-user MIMO communication is realized by $U=2$ users, each with $L=2$ RX antennas. We consider a mmWave communication of $28$GHz frequency with 5G NR OFDM waveforms of $100$MHz BandWidth (BW). Additional 5G NR waveform and system-level parameters are provided in Table.~\ref{tab:1}. The DL channels from BS to the user are assumed to be clustered mmWave channels with $100$dB pathloss. The LoS SI channel is simulated as \eqref{eq: SI_channel} with $5$mm TX-RX antenna array separation. The RX noise floors at all nodes are considered $-90$dBm, which results in $-30$dBm of RF saturation level at node $b$ for effective dynamic range of $60$dB considering $14$-bit ADCs.

Within the sensing/communication environment, $K = 4$ radar targets/scatters are considered out of which $U=2$ contributes to the DL channel. Each $K$ target is associated with a DoA $\theta_k\in[-60^{\circ}\quad 60^{\circ}],\forall k$ and a maximum range of $80$m. We have simulated $20$ radio subframes in each simulation run to evaluate the radar target estimation performance.
\begin{figure}[!tpb]
	\begin{center}
	\includegraphics[width=0.85\linewidth]{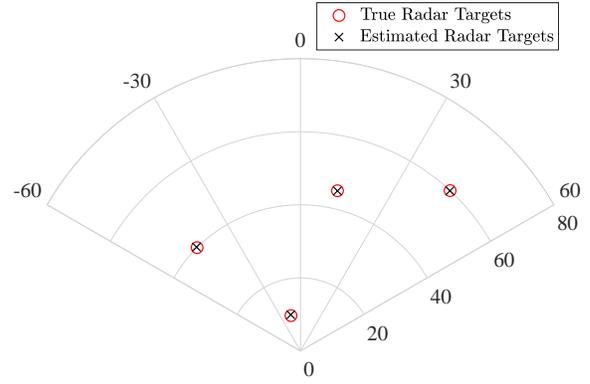}
	\caption{DoA and range estimation performance of the proposed FD ISAC system for $4$ radar targets with $N_{\rm RF} = M_{\rm RF} = 8, N_{\rm A} = N_{\rm A} = 16$  and DL transmit power of $30$dBm.}
	\label{fig: Fig2}
	\end{center}
\end{figure}

\subsection{Multiple Radar Target Sensing Performance}
In Fig.~\ref{fig: Fig2}, we have illustrated the sensing performance of the proposed multi-user FD ISAC system with a $128\times 128$ massive MIMO node transmitting DL signal with a transmit power of $30$dBm. The estimated DoA and range of each radar target are depicted in the figure. It is evident that the proposed FD ISAC approach successfully estimates the DoA and range of all the $K=4$ radar targets. The precise radar targets' DoA and range detection performance are due to the proposed target delay estimation associated with the high-resolution MUSIC approach. 

\subsection{Multi-Target Tracking Performance}
\begin{figure}[!tpb]
	\begin{center}
	\includegraphics[width=\linewidth]{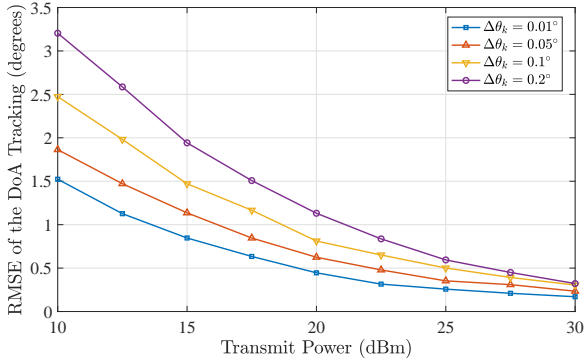}
	\caption{DoA Tracking performance for $4$ radar targets with $N_{\rm RF} = M_{\rm RF} = 8, N_{\rm A} = N_{\rm A} = 16$  and DL transmit power of $30$dBm. }
	\label{fig: Fig3}
	\end{center}
\end{figure}

In Fig.~\ref{fig: Fig3}, we have plotted the Root Mean Square Error (RMSE) of the DoA tracking in degrees for all the radar targets across $20$ 5G NR radio subframes with different DoA evolution $\Delta\theta_k = [0.01^{\circ}\,\,0.05^{\circ}\,\,0.1^{\circ}\,\,0.2^{\circ}]$ between consecutive subframes. For a range of $20$m, an angle evolution of $0.2^{\circ}$ corresponds to a velocity of $250$km/h for radar targets. Therefore, the proposed FD ISAC approach is capable of tracking multiple radar targets moving at very high speed. For a moderate transmit power of $20$dBm, the proposed approach provides less than $1^{\circ}$ RMSE for most of the DoA evolution cases. However, the RMSE of DoA tracking for all different $\Delta\theta_k$ cases is around $0.25^{\circ}$ for $30$dBm transmit power exhibiting precise target tracking performance.

\subsection{Multi-user DL Data Rate}
\begin{figure}[!tpb]
	\begin{center}
	\includegraphics[width=0.99\linewidth]{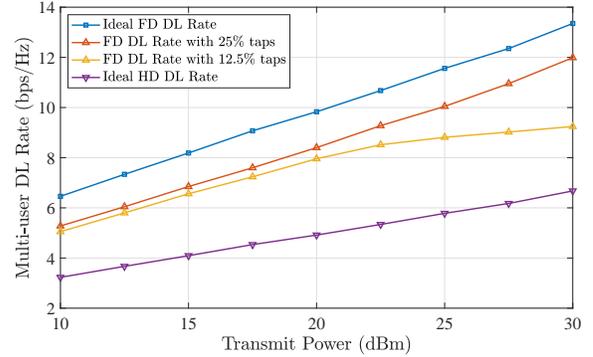}
	\caption{Multi-user DL rate w.r.t. transmit power in dBm for the $128\times 128$ massive MIMO FD BS communicating with $U=2$ DL users.}
	\label{fig: DL_rate}
	\end{center}
\end{figure}

The multi-user DL rate performance of the proposed FD ISAC system is depicted in Fig.~\ref{fig: DL_rate} for different transmit powers, where the massive MIMO node $b$ is communicating with $2$ DL users. The proposed FD ISAC approach provides the DL rate within $1$bps/Hz of the ideal FD rate up to $20$dBm transmit power. However, with the increased impact of SI at high transmit powers, the FD ISAC system with $12.5\%$ ($N=8$) SI cancellation taps exhibits rate reduction compared to the ideal FD case. In contrast, the proposed approach with $25\%$ taps provides comparable performance for all different transmit power values. Furthermore, the proposed FD ISAC system offers a superior multi-user DL rate compared to the ideal HD ISAC approach for all transmit power cases.

\section{Conclusion}
In this paper, we presented a multi-user FD ISAC framework with an FD massive MIMO node simultaneously transmitting towards multiple DL users and estimating DoA as well as range of radar targets utilizing the reflected DL waveforms. We designed a radar target tracking and DL transmission protocol across multiple communication subframes. Utilizing a limited complexity analog SI cancellation for FD massive MIMO system, we presented a joint design of the A/D beamformer and analog SI cancellation that maximizes both radar target tracking and multi-user DL rate performance. Our performance results considering a mmWave channel model exhibited the high precision DoA and rage estimation of multiple radar targets while providing maximized multi-user DL rate.

\section*{Acknowledgments}
This work was partially funded by the National Science Foundation CAREER award \#1620902.

\bibliographystyle{IEEEtran}
\bibliography{IEEEabrv,ms}

\end{document}

%% file: Definitions.tex
\renewcommand{\a}{\mathbf{a}}

\newcommand{\g}{\mathbf{g}}

\newcommand{\n}{\mathbf{n}}

\renewcommand{\r}{\mathbf{r}}
\newcommand{\s}{\mathbf{s}}
\renewcommand{\t}{\mathbf{t}}

\renewcommand{\v}{\mathbf{v}}
\newcommand{\w}{\mathbf{w}}
\newcommand{\x}{\mathbf{x}}
\newcommand{\y}{\mathbf{y}}
\newcommand{\z}{\mathbf{z}}

\newcommand{\B}{\mathbf{B}}
\newcommand{\C}{\mathbf{C}}
\newcommand{\D}{\mathbf{D}}
\newcommand{\E}{\mathbf{E}}
\newcommand{\F}{\mathbf{F}}
\newcommand{\G}{\mathbf{G}}
\renewcommand{\H}{\mathbf{H}}
\newcommand{\I}{\mathbf{I}}

\newcommand{\R}{\mathbf{R}}

\newcommand{\U}{\mathbf{U}}
\newcommand{\V}{\mathbf{V}}
\newcommand{\W}{\mathbf{W}}

\newcommand{\Compl}{\mbox{$\mathbb{C}$}}
